\documentclass[twocolumn,english,prb,showpacs]{revtex4}
\usepackage[latin9]{inputenc}
\usepackage{amssymb}
\usepackage{graphicx}
\usepackage{amsmath}
\usepackage{color}
\usepackage{mathrsfs}
\usepackage{float}
\usepackage{indentfirst}
\usepackage{mathrsfs}
\usepackage{float}
\usepackage{indentfirst}
\usepackage{textcomp}
\usepackage{comment}
\usepackage{mathtools}

\newcommand{\COMMENTED}[1]{}

\begin{document}

\title{Coupling quantum Monte Carlo and independent-particle calculations: \\
self-consistent constraint for the sign problem 
based on density or density matrix}

\author{Mingpu Qin}
\affiliation{Department of Physics, College of William and Mary, Williamsburg, Virginia 23187}

\author{Hao Shi}
\affiliation{Department of Physics, College of William and Mary, Williamsburg, Virginia 23187}

\author{Shiwei Zhang}
\affiliation{Department of Physics, College of William and Mary, Williamsburg, Virginia 23187}

\begin{abstract}
	Quantum Monte Carlo (QMC) methods are one of the most important tools for studying
	interacting quantum many-body systems.
The vast majority of QMC calculations in interacting fermion systems require a 
constraint to control the sign problem. The constraint involves an input trial wave function 
which restricts the random walks. We introduce a systematically improvable constraint
which relies on the fundamental role of the density or one-body density matrix.
An independent-particle calculation is coupled to an auxiliary-field QMC calculation.
The independent-particle solution is used as the constraint in QMC, 
which then produces the input density or density matrix for the next iteration.
The constraint is optimized by the self-consistency between the 
many-body and  independent-particle calculations. The approach is demonstrated in the two-dimensional 
Hubbard model by accurately determining the 
ground state when 
collective modes separated by tiny energy scales are present 
in the magnetic and charge correlations.
Our approach also provides an \emph{ab initio} way to predict effective 
interaction parameters for 
 independent-particle calculations.  
\end{abstract}


\maketitle

\section{introduction}
The study of interacting quantum many-body systems presents a major challenge in modern
physics.
Quantum Monte Carlo (QMC) methods \cite{QMC_prd_1981,QMC_AP_1986,QMC_prb_1989,QMC_rmp_2011} are a
key numerical approach for solving such systems.
The dimension of the Hilbert space involved in a quantum many-body system grows exponentially with the system 
size. QMC methods can in principle provide stochastic evaluations of expectation values in such 
systems with computer times that scale polynomially with system size.
However, with a few exceptions \cite{Hirsch_prb_1985, Wu_prb_2005}, 
direct QMC calculations  in fermion systems 
suffer from the minus sign problem \cite{sign_1,sign_2}, 
which breaks this scaling.
The most effective approach for dealing with the sign problem in general has been by a bias-variance trade-off.
A constraint is applied in \emph{some} space to restrict the Monte Carlo sampling,
which introduces a systematic bias but in turn removes the exponential growth in variance and restores
the algebraic complexity of the algorithm. The majority of QMC calculations have employed this approach,
including many on spin and fermion models \cite{GF_fermion,paper_simons}, and almost all on realistic systems
in condensed matter physics \cite{DMC_RMP,DMC_Devaux,AFQMC-solids}, nuclear physics \cite{CP_nuclear}, and quantum chemistry \cite{DMC_QC_1,DMC_QC_2,AFQMC-QC}.

 A missing link in such an approach is that it has been difficult to make the constrained QMC calculations
 systematically improvable without drastically changing its computational scaling or complexity \cite{diff_self}. 
  Although the calculations are often among the most accurate possible for many-fermion
  systems \cite{paper_simons}, the accuracy cannot be assessed internally, and 
 there has not been a conceptual framework which allows one 
 to build on the outcome of the calculation in a practical way to further
 reduce the systematic error from the constraint. The constraint typically relies on a trial wave 
 function which is provided by an external source (e.g., an independent-electron calculation or 
 a variational Monte Carlo optimization \cite{VMC}), and a ``one-shot'' answer is obtained from the QMC. 
 
 In this paper we introduce a self-consistent constraint in QMC using the auxiliary-field QMC (AFQMC) framework.
 The approach couples the AFQMC calculation to an independent-electron calculation which provides the 
 trial wave function for the constraint. The spin densities (or density matrix) obtained from the QMC
 are then fed back into the independent-electron calculation, whose effective interaction 
 strength (or more generally, exchange-correlation functional) is tuned to best match the QMC densities.
 The output wave function is then used for a new AFQMC calculation, and the process
 is iterated to convergence. We show that this procedure allows the calculations to systematically
 improve. 
 The QMC can recover from an initial constraint in a wrong state,  
 i.e., one with an incorrect 
 magnetic order, and provides the correct prediction at convergence even when a small residual 
 constraint error is still present.
 
 In an alternative, complementary view, the self-consistent approach is motivated by the 
 fundamental role of the electron density or density matrix in many-fermion systems \cite{Gilbert_1975}. 
 The  constraining wave function in AFQMC is usually taken as a
 solution from the Hartree-Fock (HF) or a density-functional theory (DFT) 
 calculation based on the same many-body Hamiltonian. 
 What is the optimal independent-electron wave function? 
 Our approach defines
 a procedure for determining the answer. We will show that, by varying the strength of the Coulomb 
 repulsion, the HF
 calculation can give
 better order parameters (spin densities here). The self-consistency procedure with QMC
 thus provides an optimal effective ``$U$'' parameter. This can potentially be used to 
 derive effective Hamiltonians
 to be used by less computing-intensive methods in larger system sizes.
 Similarly, in the context of DFT, the procedure would define a way to find an optimal functional
 (in the spirit of hybrid functionals, for example).
 
 For concreteness, we will use the Hubbard model to describe the self-consistent AFQMC procedure:  
\begin{equation}
\hat H=-\sum_{i,j;\sigma} t_{ij} c_{i\sigma}^{\dagger}c_{j\sigma}+\sum_{i} U{\hat n_{i\uparrow}}{\hat n_{i\downarrow}}+\sum_{i} v_{i,\sigma} {\hat n_{i\sigma}}\,,
\label{eqn:H}
\end{equation}
where $c_{i\sigma}^\dagger(c_{i\sigma})$ is the creation (annihilation) operator on lattice site $i$,
$\sigma = \uparrow,\downarrow$ is the spin of the electron, $\hat n_{i,\sigma}=c_{i\sigma}^\dagger c_{i\sigma}$ 
is the number operator. The hopping matrix elements $t_{ij}$, on-site interaction strength $U$, and spin-dependent
external potential $v_{i, \sigma}$ are parameters. 
The overall electron density is given by the parameter 
$n \equiv (N_{\uparrow}+N_{\downarrow})/{N}$, with $N$ being the total number of lattice sites, and the 
hole density is then $h=1-n$. 

\section{Self-consistent method coupling with independent-electron calculations}
The corresponding independent-particle (IP) calculation treats a Hamiltonian of the form:
 \begin{equation}
\hat H_{\rm IP}^{\sigma} =  -\sum_{i,j} t_{ij} c_{i\sigma}^{\dagger}c_{j\sigma} 
+\sum_{i}U_{\rm eff}\langle \hat n_{i\bar{\sigma}}\rangle {\hat n_{i\sigma}}
+\sum_{i} v_{i\sigma} \hat n_{i\sigma}\,,
\label{mf_eq_up}
\end{equation} 
where $\bar{\sigma}$ denotes the opposite of $\sigma$. In the standard unrestricted HF (UHF) calculation,
$U_{\rm eff}$ takes the ``bare'' value of $U$, and 
the input mean-field is a set of expectation values, $\langle \hat n_{i\bar{\sigma}}\rangle$, 
computed with respect to the 
solution from the previous IP step. In DFT with a local spin-density type of
approach, $U_{\rm eff} \langle \hat n_{i\bar{\sigma}}\rangle$ in Eq.~(\ref{mf_eq_up}) is replaced by
an exchange-correlation functional, $V_{\rm xc}[ \langle \hat n_{i\sigma}\rangle]$.
The wave function from the IP solution is a single Slater determinant, 
$|\psi\rangle = |\psi_\uparrow\rangle \otimes |\psi_\downarrow\rangle$, with
$|\psi_\sigma\rangle = \phi_1^\dagger \phi_2^\dagger \cdots \phi_{N_\sigma}^\dagger |0\rangle$
where $\phi_{i}^\dagger = \sum_{j} \phi^\sigma_{j i} c^\dagger_j$ creates 
a $\sigma$-spin electron in the single-particle orbital given by the vector $\{\phi^\sigma_{j i},j=1, \cdots, N \}$.

The AFQMC method 
projects the many-body ground state
of ${\hat H}$ in Eq.~(\ref{eqn:H}) by an iterative process:
$\lim_{m                                                                                  
\to \infty} (e^{-\tau {\hat H}})^m \left| \psi_T \right\rangle \propto                    
\left| \Psi_0 \right\rangle$, 
where $\tau>0$ is a small parameter.
For convenience, we take the initial 
state, which must be non-orthogonal to $\left| \Psi_0 \right\rangle$, 
 to be a single Slater determinant trial wave function, $\left| \psi_T \right\rangle$.
 The many-body propagator is written as 
 $e^{-\tau \hat H}\doteq
\int p(x)\,e^{\hat h(x)}\,dx$ 
where $\hat h(x)$ is a general IP
``Hamiltonian''
dependent on the multi-dimensional
vector   $x$, and $p(x)$ is a probability density function \cite{lecture-notes}.
The interacting many-body
system is thus mapped into
a linear combination of many IP systems in 
fluctuating auxiliary fields, $x$. 
The AFQMC method represents the many-body wave function 
 as an ensemble of Slater determinants, i.e., $|\Psi_0\rangle=\sum_{k}\omega_{k}|\psi_{k}\rangle$. 
The iterative projection is realized by a random walk in Slater determinant space, 
in which for each walker $|\psi_{k}\rangle$, an auxiliary field $x$ is sampled from $p(x)$, 
and the walker is propagated: $e^{\hat h(x)} |\psi_k\rangle \rightarrow |\psi_k'\rangle$.
Computationally this  is similar to a step in the IP calculation.

Because the propagator $e^{\hat h(x)}$ contains stochastically fluctuating fields, 
the random walks 
will, except for special cases protected by symmetry \cite{Hirsch_prb_1985}, 
reach Slater determinants with arbitrary sign or phase \cite{lecture-notes}. In representing the 
ground state, only one from \emph{each pair} of Slater determinants $\{\pm|\psi\rangle\}$ 
(or from
\emph{the set} $\{e^{i\theta} |\psi\rangle\}$) is needed. When both (all) are present in the samples, 
the wave function signal is lost in noise, 
because the Monte Carlo  weights, $w_k$, are always positive.
This is the sign (phase) problem \cite{lecture-notes}.
For the Hubbard Hamiltonian, $\hat h(x)$ is real, so ``only'' a sign problem appears. 
To control this problem, 
we use the 
trial wave-function $|\psi_T\rangle$ for importance sampling, which
guides the random walk and constrains it to only half of the Slater determinant space: $\langle \psi_T|\psi_k\rangle >0$  \cite{zhang_prb_1997}.
This approach has been referred to as the constrained-path Monte Carlo (CPMC) method. 
For a general Hamiltonian with two-body interactions,  
a generalized gauge condition allows a similar framework for the phase problem \cite{zhang_prl_2003, lecture-notes}.

This framework eliminates the sign or phase problem, at the cost of introducing a systematic bias.
Previous studies 
in a variety of systems have 
shown that the bias tends to be small, in both models \cite{paper_simons,chia-chen_prb, hubbard_benchmark,CPMC_sym_1,CPMC_sym_2} 
and realistic materials \cite{AFQMC-QC,lecture-notes,zhang_prl_2003,AFQMC-solids},
making this one of the most accurate many-body approaches for general interacting 
fermion systems.
In this work, we introduce a self-consistent method to further reduce the bias introduced by the constraint
from the trial wave function.


To start the self-consistent procedure, we first carry out a CPMC calculation for the many-body Hamiltonian,
Eq.~(\ref{eqn:H}), using any typical choice of $|\psi_T\rangle$, for example a non-interacting wave function or 
the 
UHF solution. We use back-propagation 
\cite{zhang_prb_1997,Wirawan-PRE} 
 to compute the expectation values of the quantities that do not commute with $\hat H$. 
 After the CPMC calculation, we solve the IP Hamiltonian in Eq.~(\ref{mf_eq_up}),
 using the densities obtained from the preceding QMC calculation as the input mean field,
 i.e., $\langle \hat n_{i\bar\sigma} \rangle_{\rm QMC} \rightarrow \langle \hat n_{i\bar \sigma} \rangle$.  
An effective interaction, $U_{\rm eff}$, is applied  instead of the ``bare'' $U$ value.
 We vary  $U_{\rm eff}$ to find an optimal value whose solution gives densities closest 
 to the input from QMC, i.e., the $U_{\rm eff}$ which minimizes:
 \begin{equation}
\delta=\frac{1}{N} \big(\sum_{i\sigma}(\langle \hat n_{i\sigma}\rangle_{\rm IP}-\langle \hat n_{i\sigma}\rangle_{\rm QMC})^{2}\big)^{1/2}\,.
\label{delta_def}
\end{equation}
The IP solution with the optimal  
$U_{\rm eff}$ determined from Eq.~(\ref{delta_def}) is then used as the $|\psi_T\rangle$ in a new 
 CPMC calculation. This procedure is iterated until convergence. 

\section{Results}

We use the two-dimensional Hubbard model at density $n=0.875$ as a test case.
The hopping matrix element $t_{ij}$ is $t$ for nearest neighbors and 0 otherwise, and we set the interaction $U=8t$.
This parameter regime, mimicking the situation in doped cuprates \cite{stripe_exp}, is notoriously challenging,
and its ground state in the thermodynamic limit remains unknown.
We focus on the nature of the magnetic and charge correlations, which is crucial for
understanding the properties of lightly doped antiferromagnets.

\COMMENTED{
and scan lattice sizes to 
accommodate the
wavelength of possible collective modes \cite{chia-chen_prl} of the  underlying antiferromagnetic (AFM) order. 
Using the external potential $v_{i\sigma}$, we apply pinning fields \cite{srwhite_prl_2007,srwhite_prb_2009,assaad_prx_2013} 
in localized regions to break translational symmetry. In this way, two-body 
spin and charge correlation functions are turned into one-body spin and charge order parameters, which
are simpler to measure in the calculation.
Thus, aside from the total energy,  we will focus on the local spin and charge densities
  \begin{equation}
  h(i) = 1 - \langle \hat n_{i\uparrow} + \hat n_{i\downarrow}\rangle; \ \ 
 S_z(i) =  \langle \hat n_{i\uparrow} - \hat n_{i\downarrow}\rangle/2\,.
 \label{S-z_h_def}
\end{equation}
}

In our test calculations, we will
consider a cylindrical geometry, i.e., cells with periodic boundary conditions
in the  $x$-direction and open boundary conditions in the $y$-direction, 
for which density matrix renormalization group (DMRG) \cite{sr_white_dmrg_original} 
calculations can give very accurate benchmark results for systems of 
significant sizes. 
 We denote the integer coordinates of the lattice site $i$ by $(i_x,i_y)$. 
 Local antiferromagnetic (AFM) correlations are expected, and our calculations aim to 
 probe what happens to the AFM correlation in this regime of doping ($h=1/8$) and  
 strong interaction ($U=8t$).
 To break degeneracy from translational symmetry,
 pinning fields \cite{srwhite_prl_2007,srwhite_prb_2009,assaad_prx_2013} 
 are applied at the edges: 
  $v_{i\uparrow}=-v_{i\downarrow}=(-1)^{i_x}\nu_0$ 
 for $i_y=1$ and $i_y=L_y$. 
 We set the  pinning field strength $\nu_0=t/4$ in all calculations.
  With pinning fields, two-body 
spin and charge correlation functions in periodic systems are turned into one-body spin and charge order parameters, which are simpler to measure in the calculation.
Thus, in addition to the total energy,  we will focus on the local spin and charge densities
  \begin{equation}
  h(i) = 1 - \langle \hat n_{i\uparrow} + \hat n_{i\downarrow}\rangle; \ \ 
 S_z(i) =  \langle \hat n_{i\uparrow} - \hat n_{i\downarrow}\rangle/2\,.
 \label{S-z_h_def}
\end{equation}

 \begin{figure}[t]
 	\includegraphics[width=4.26cm]{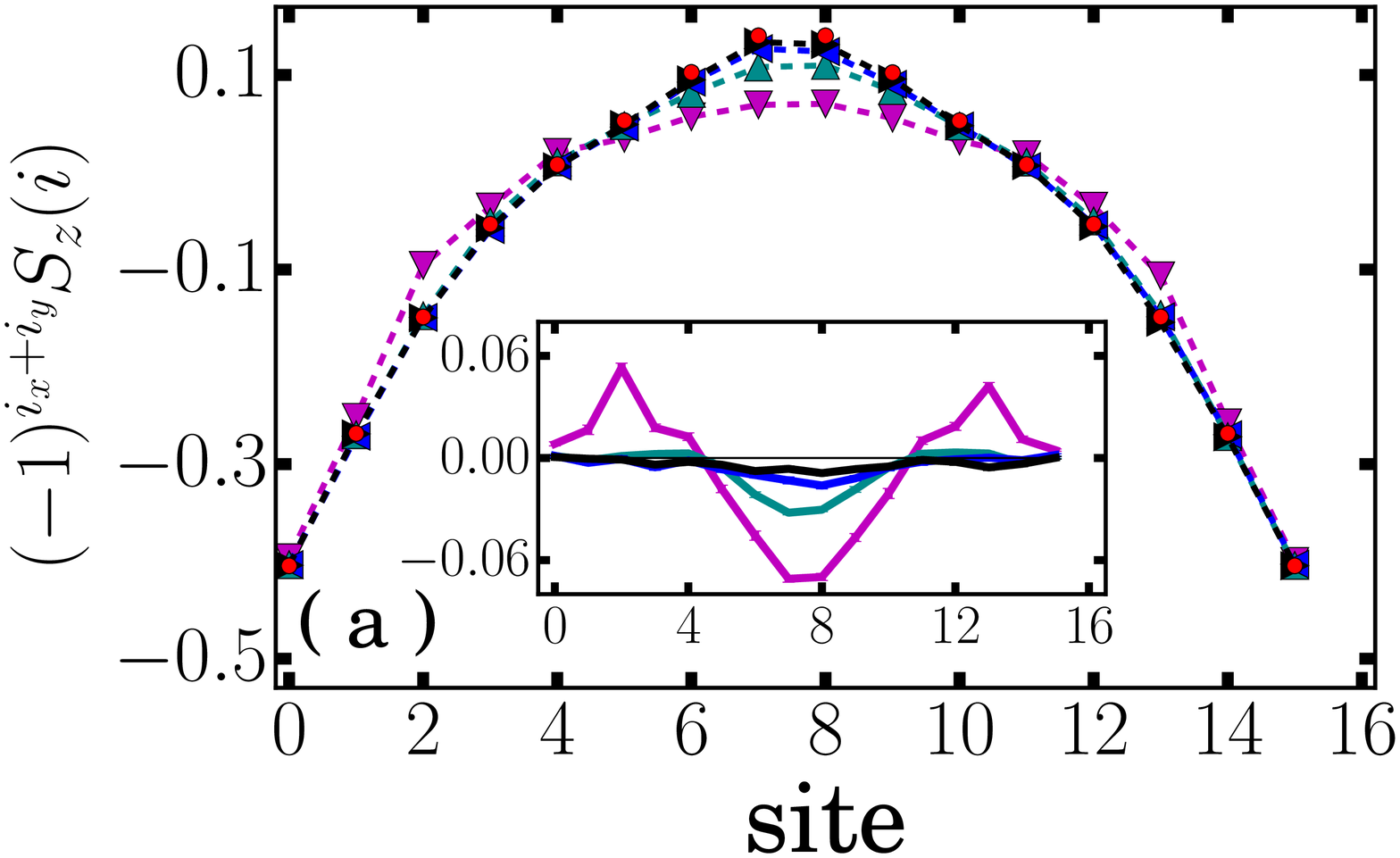}
 	\includegraphics[width=4.0cm]{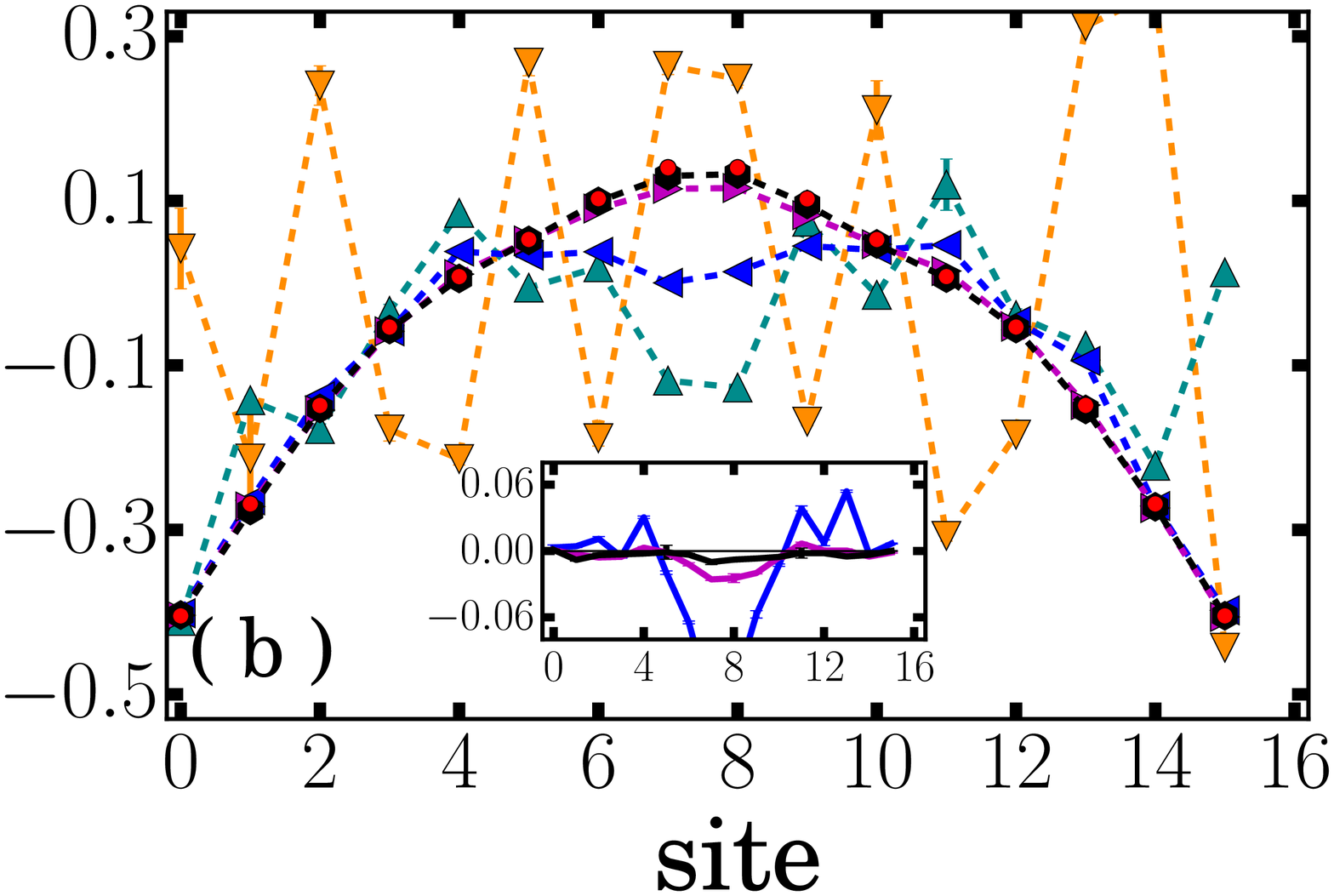}
 	\includegraphics[width=4.26cm]{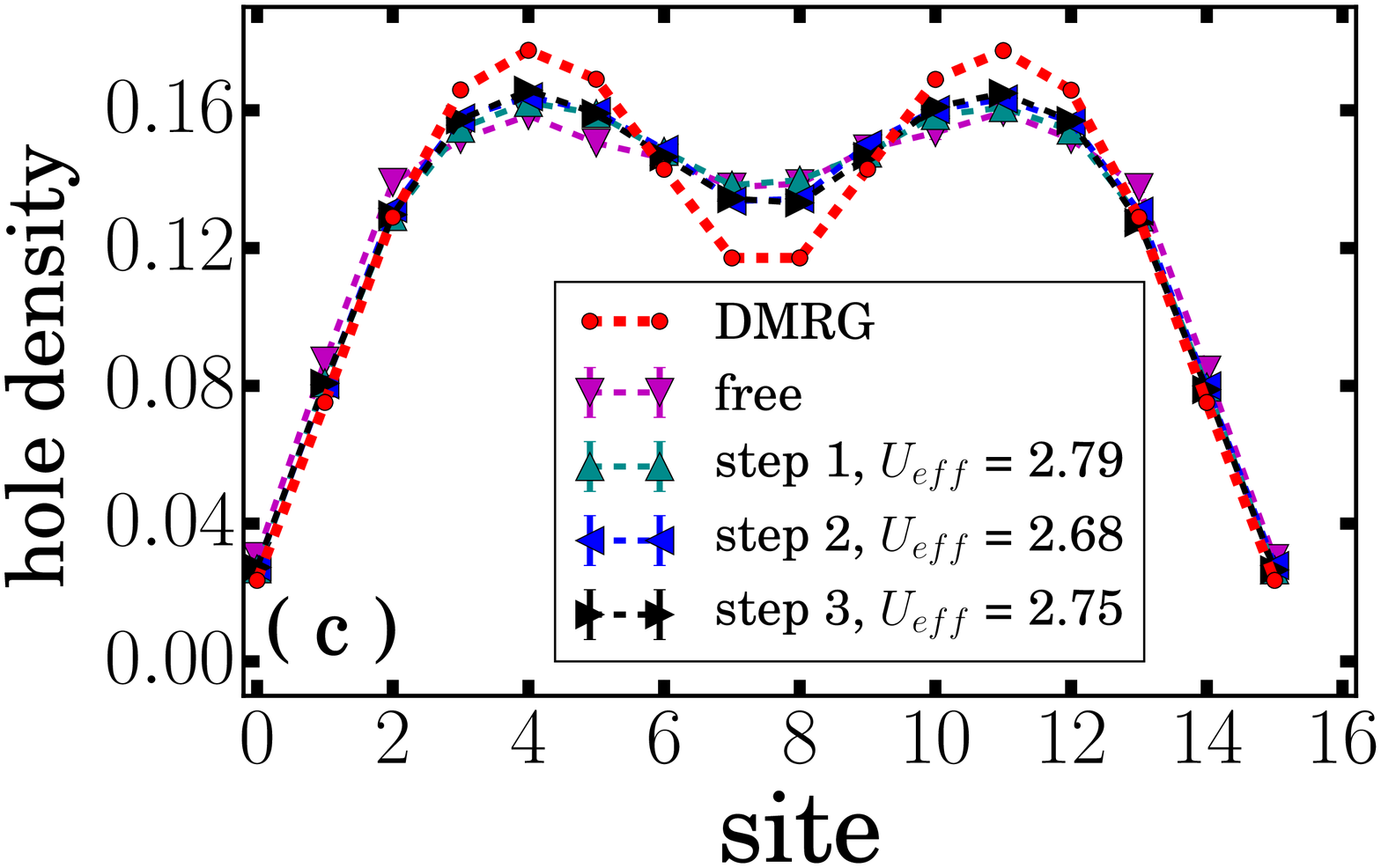}
 	\includegraphics[width=4.0cm]{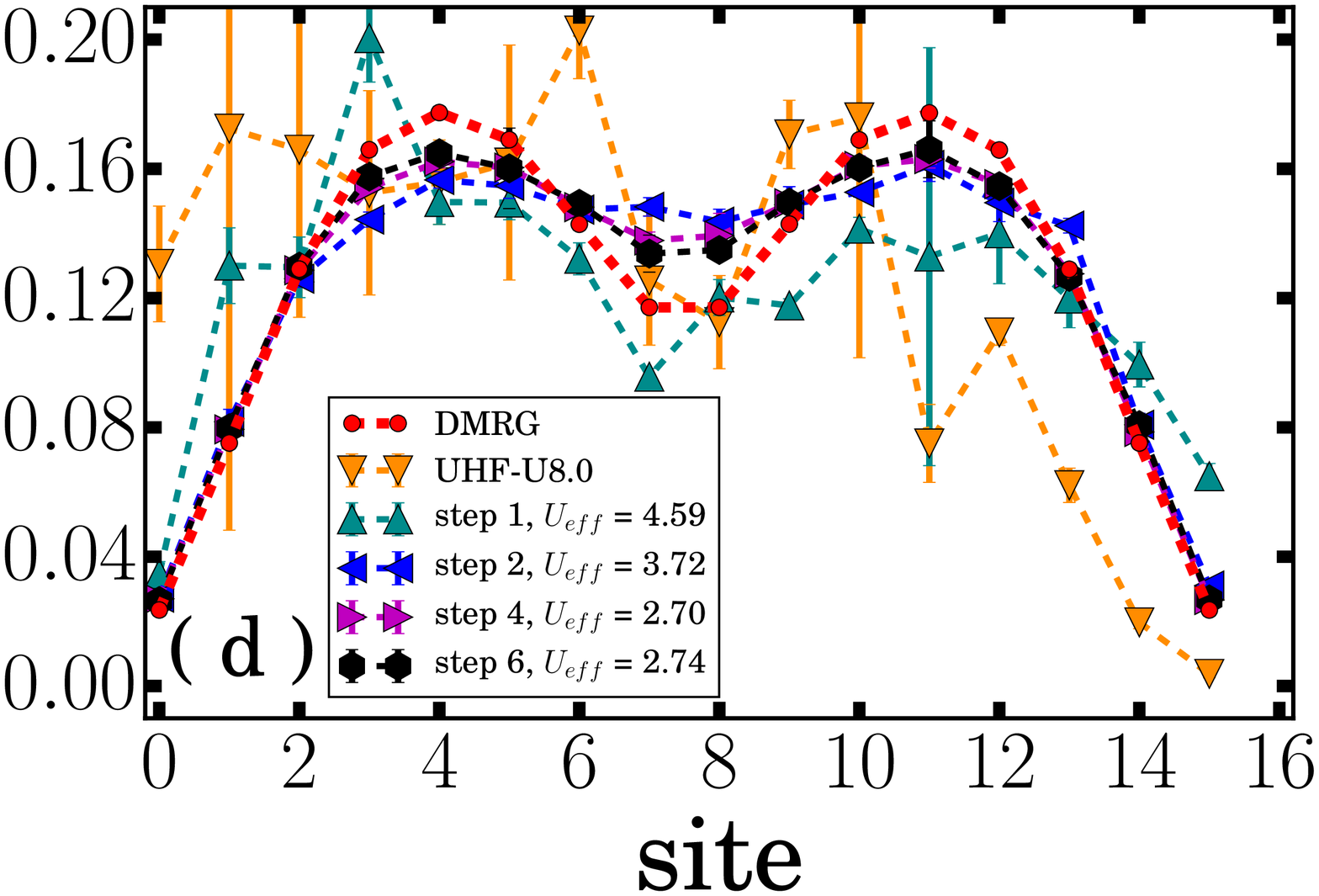}
 	\caption{Systematic improvement of the CPMC accuracy
	from the self-consistent procedure.
	The top panel plots staggered spin density along the $y$-direction vs.~site label $i_y$ ($i_x=1$). 
	The bottom panel plots the corresponding hole density. 
	The left and right columns show
	two self-consistent procedures starting from two different initial $|\Psi_T\rangle$'s, the free-electron 
	wave function and the UHF solution at $U=8t$, respectively.	
	In the legend, the $U_{\rm eff}$ value of the IP calculation is listed for each
	iteration step.
	In (a) and (b), the differences  
	 at different stages of the  iteration, with respect to the reference DMRG results, are shown in the insets.
	The system is $4 \times 16$, with $U = 8t$, $h=1/8$ doping, with pinning field applied to both edges along $L_y$. 
 	}
 	\label{spin_4_16_mf}
 \end{figure}

We first illustrate the method  in a $4 \times 16$ system.
Two different choices of the trial wave function $|\psi_T\rangle$ are used for the initial CPMC calculation.
The first is the ground state of the corresponding non-interacting Hamiltonian (referred to as free-electron hereafter).
The other is the UHF 
solution \cite{xu_jpcm_2011} obtained with
the ``bare'' $U$ value, i.e., $U = 8t$. 
In Fig.~\ref{spin_4_16_mf}, we show the staggered spin densities, $(-1)^{i_{x}+i_{y}}S_z(i)$,
and the hole densities $h(i)$ computed by the self-consistent QMC procedure, and compare them 
with DMRG results, which are essentially exact for this system.  
The results are, as expected,  statistically invariant with respect to $i_x$,  and are only shown for
$i_x=1$ \cite{diff_row}.
The staggered spin densities, shown in the upper panel, depict a modulation of the AFM order. (The magnetic
moments are the strongest at the edges because of the pinning fields.) 
At a node when the curve crosses zero, a $\pi$ phase shift is created in the AFM pattern. 
The holes tend to concentrate at the nodes, creating the peaks seen in the bottom panel. 

We see from  Fig.~\ref{spin_4_16_mf}  that, independent of which $|\psi_T\rangle$ is used in the initial CPMC
calculations, the self-consistent procedure leads to a systematic improvement of the spin densities which 
approach the DMRG results. Convergence from the free-electron $|\psi_T\rangle$ requires only about 
$3$ iterations between the QMC and IP calculations. 
The UHF initial $|\psi_T\rangle$, which predicts a wrong phase (see below),  gives results in the 
first-iteration CPMC with large errors. The self-consistency quickly recovers and converges in 
about $6$ iterations. 
The difference between converged CPMC spin densities and those from DMRG, as seen in the insets, is very 
small. The hole density, which shows a slightly larger 
residual error, clearly gives the correct charge pattern.
 The self-consistent CPMC method thus accurately determines the 
 ground state and its magnetic order in this system.

 \begin{figure}[t]
 	\includegraphics[width=4.08cm]{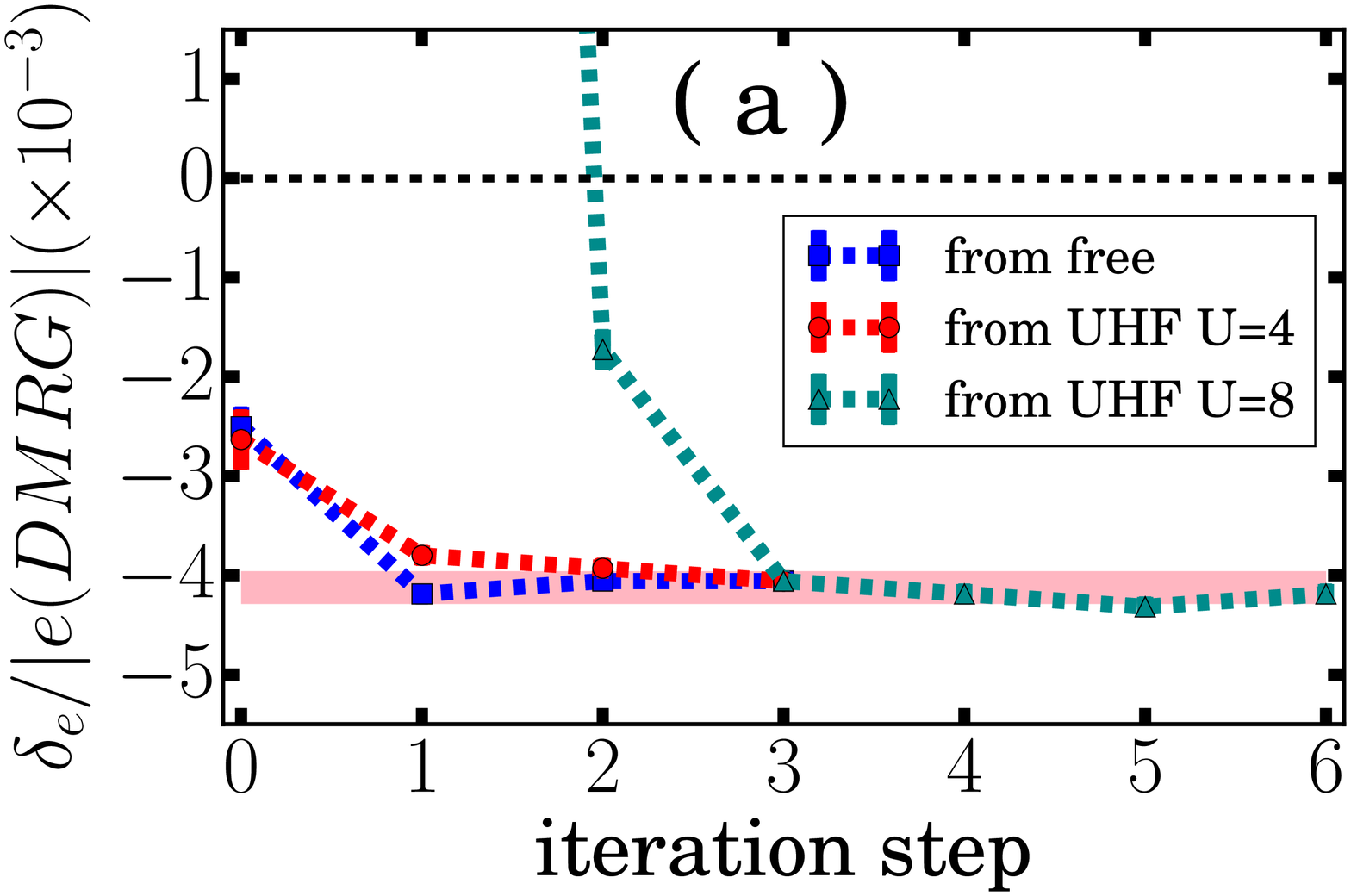}
 	\includegraphics[width=4.0cm]{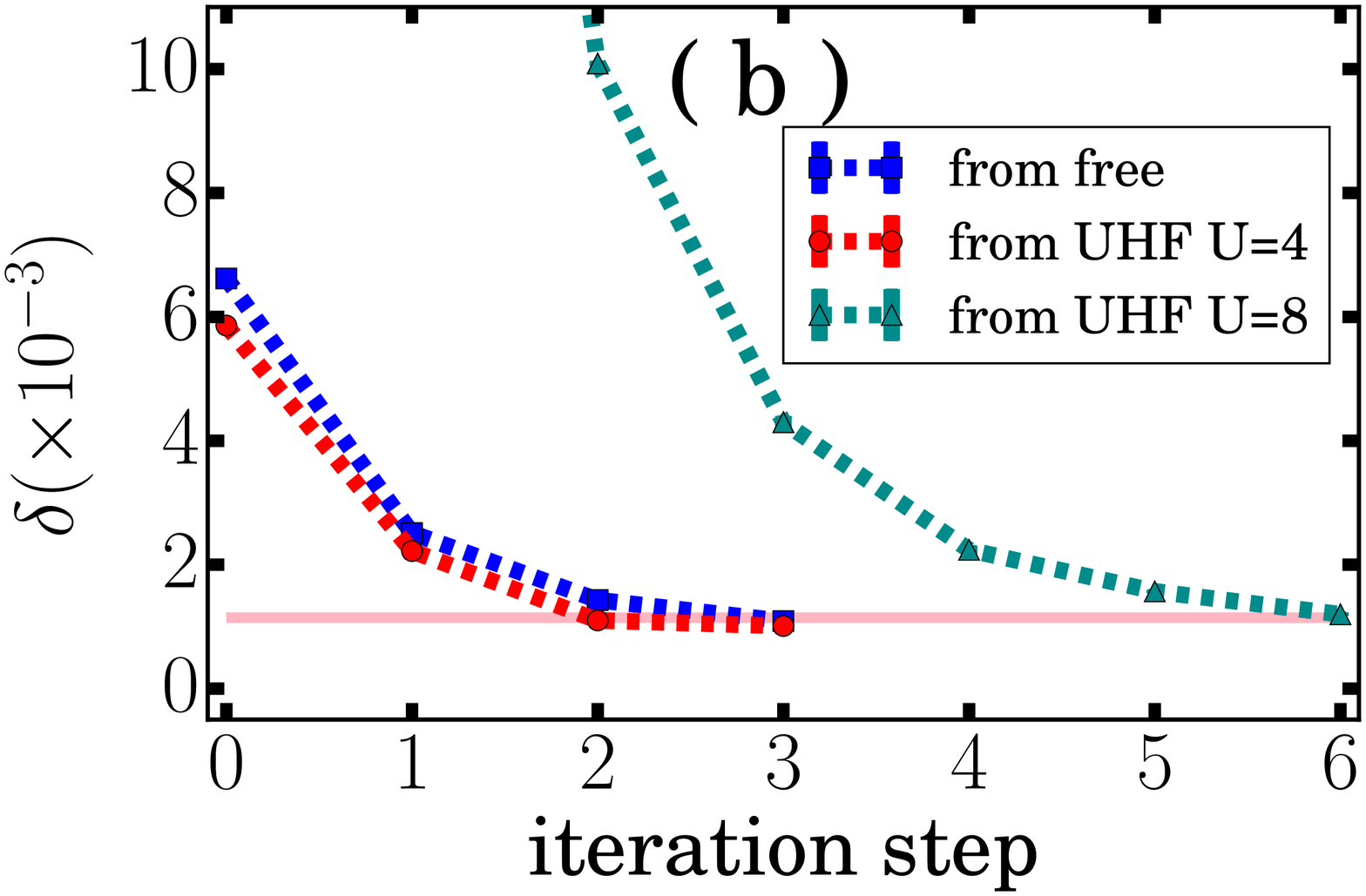}
 	\caption{Convergence of CPMC results in the self-consistent procedure.
	Three initial trial
 		wave-functions are tested: free-electron and UHF solutions at $U = 4t$ and $8t$.
		The system is the same as in Fig.~\ref{spin_4_16_mf}.
		In each panel, the pink band represents the converged result and error bar.
	In (a), the relative error in the computed ground-state energy relative to DMRG is shown vs.~self-consistency iteration. The horizontal dotted line at zero is to aid the eye. 
	Plotted in (b) are the mean square error [following the definition in Eq.~(\ref{delta_def})] in the local density computed by CPMC from the reference DMRG results.} 
 	\label{e_mf}
 \end{figure}

The convergence process of the self-consistent procedure is further illustrated in Fig.~\ref{e_mf}.
The left panel shows the relative error in the CPMC ground-state energy per site, from the reference DMRG value of   
 $-0.77127(2)$.
The energy is seen to converge, independent of the initial $|\Psi_T\rangle$,
to a value with a residual
relative error about $-0.4 \%$.
We note that the mixed estimate, which is used in CPMC to computed the energy, is not variational \cite{zhang_prb_1997,Carlson-PRB1999}.
The self-consistency actually leads in a 
slightly worse ground-state energy than the initial CPMC results 
computed with the free-electron or UHF solution with $U_{\rm eff}=4t$ as $|\Psi_T\rangle$.
On the other hand, the $U = 8t$ 
UHF solution gives an incorrect state with wrong AFM order \cite{xu_jpcm_2011}. 
Using it as $|\Psi_T\rangle$ in a one-shot CPMC calculation, the   
relative error of ground state energy 
is $\sim 11 \%$. Clearly 
the self-consistent process leads to a large improvement.

\begin{figure}[t]
 	\includegraphics[width=4.26cm]{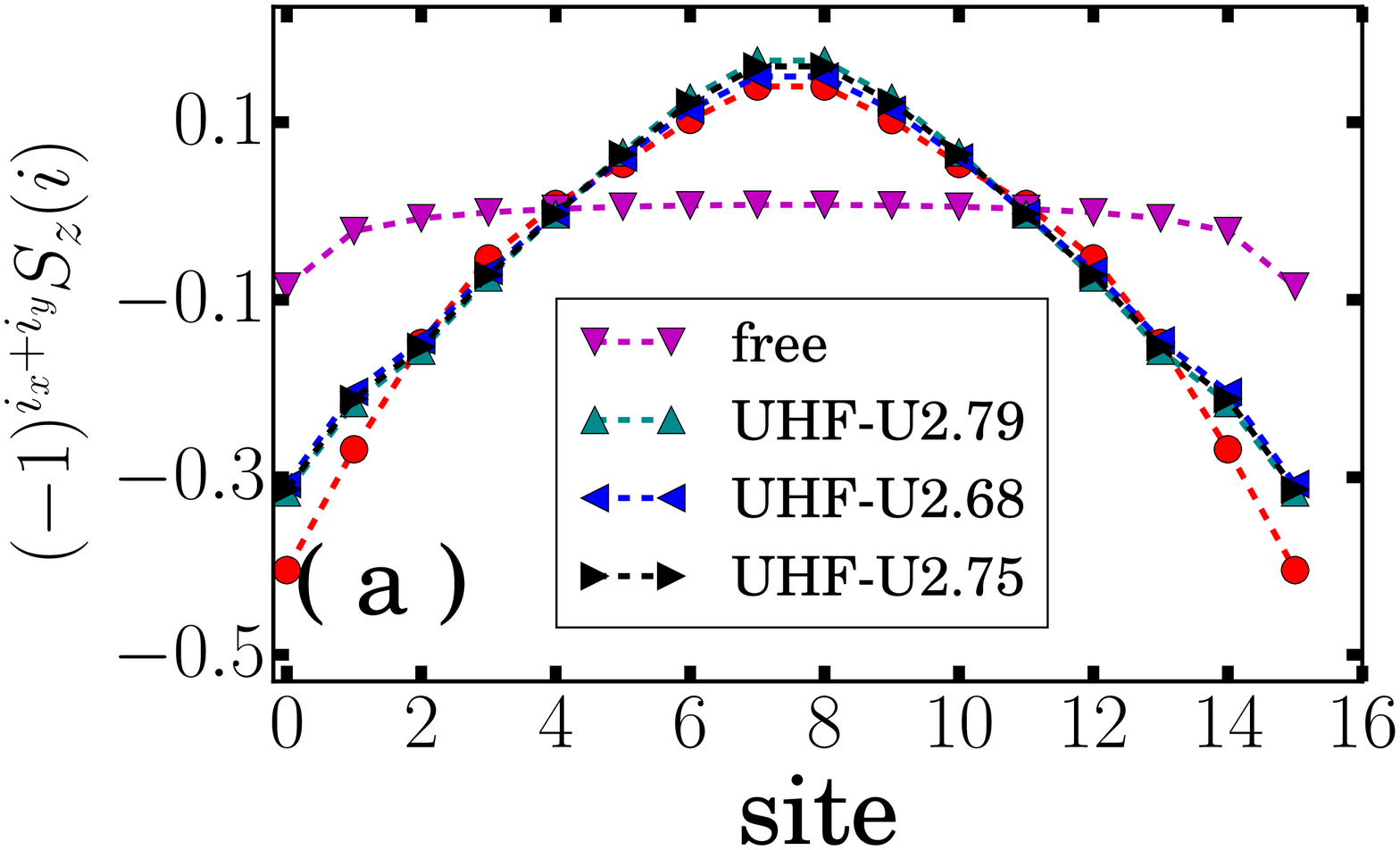}
 	\includegraphics[width=4.0cm]{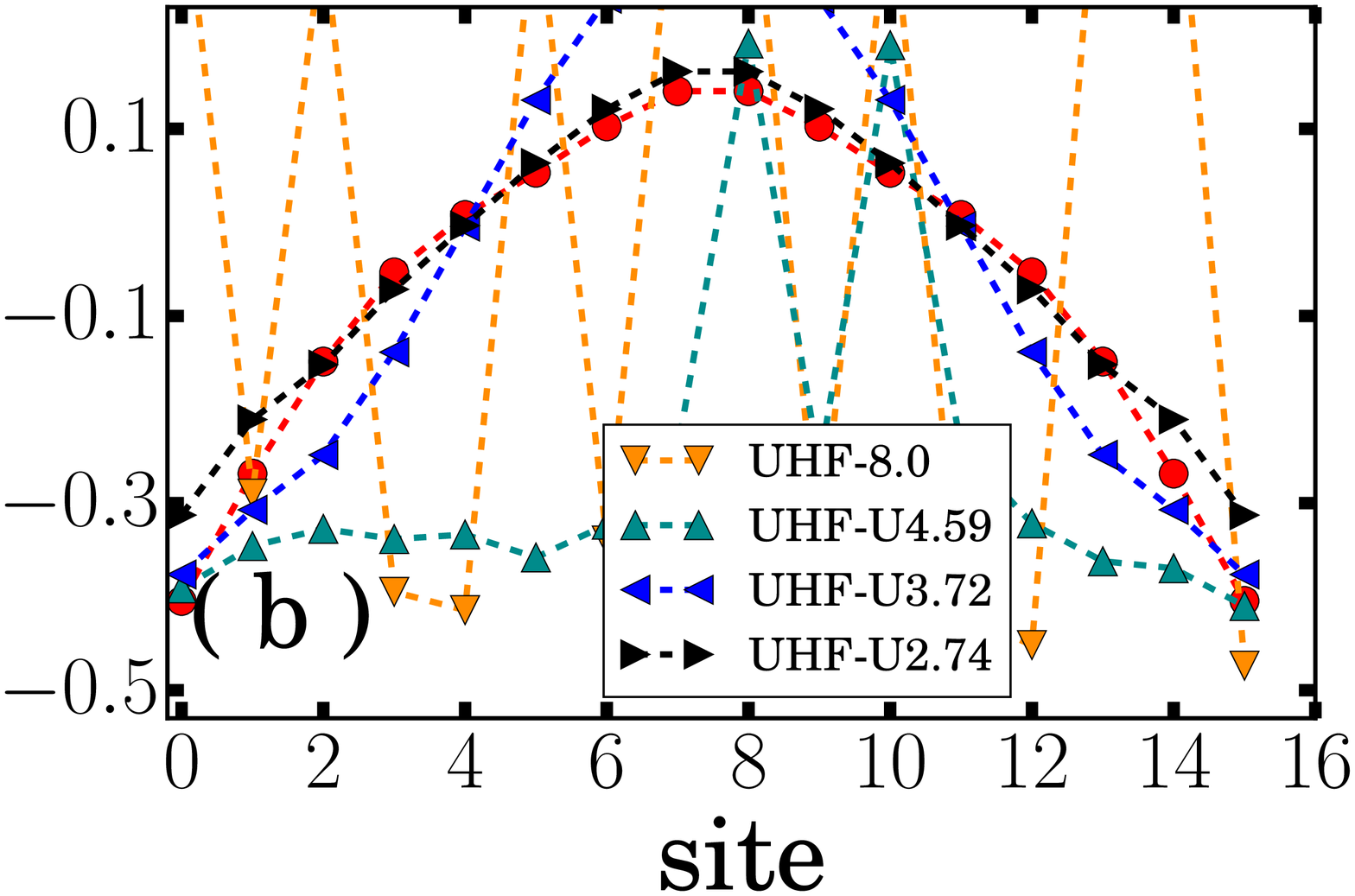}
 	\includegraphics[height=0.68cm]{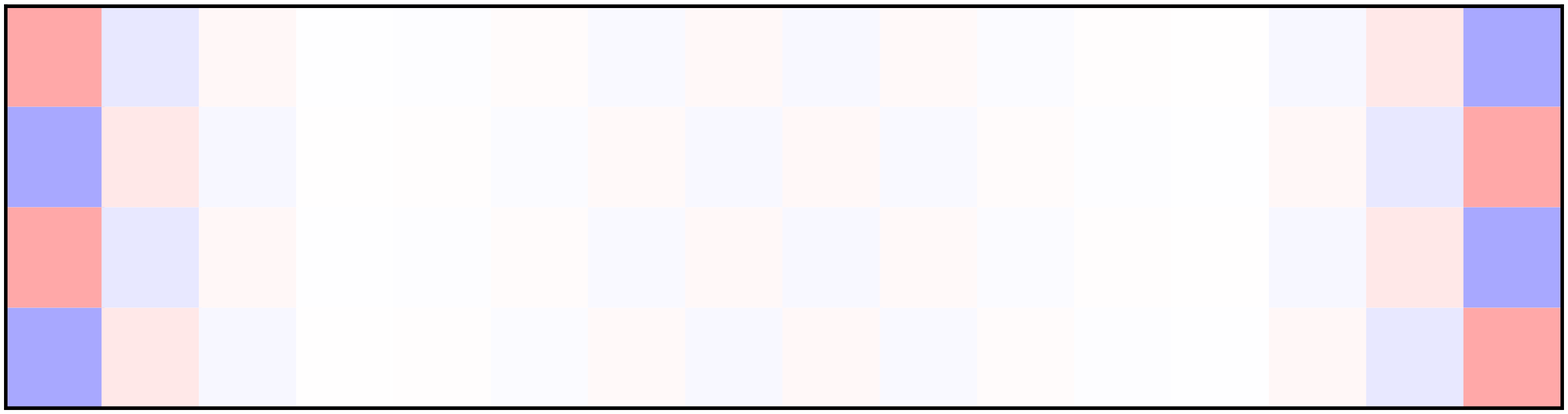}
 	\includegraphics[height=0.68cm]{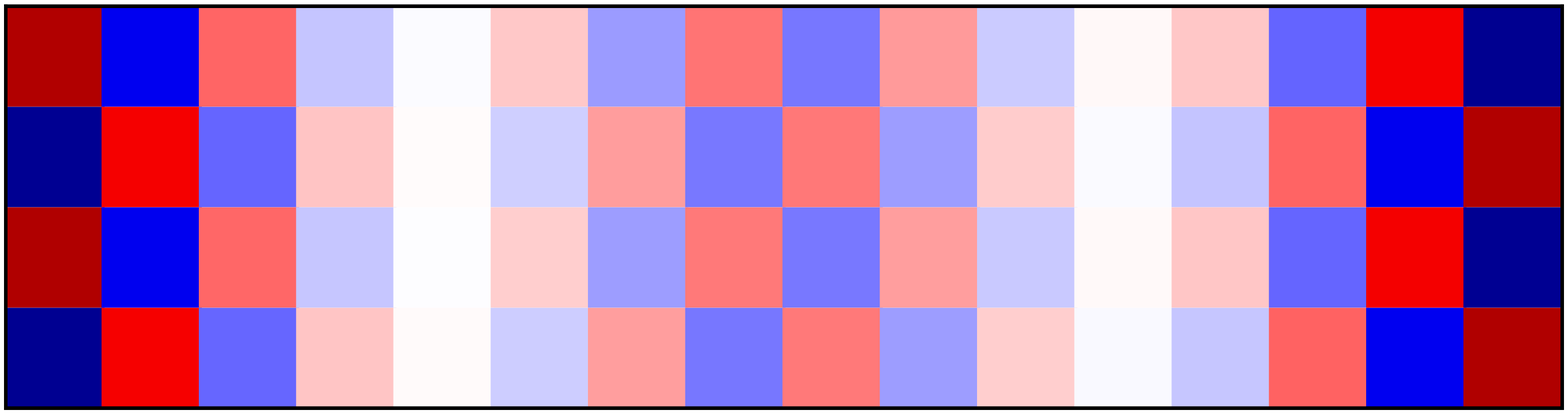}
 	\includegraphics[height=0.68cm]{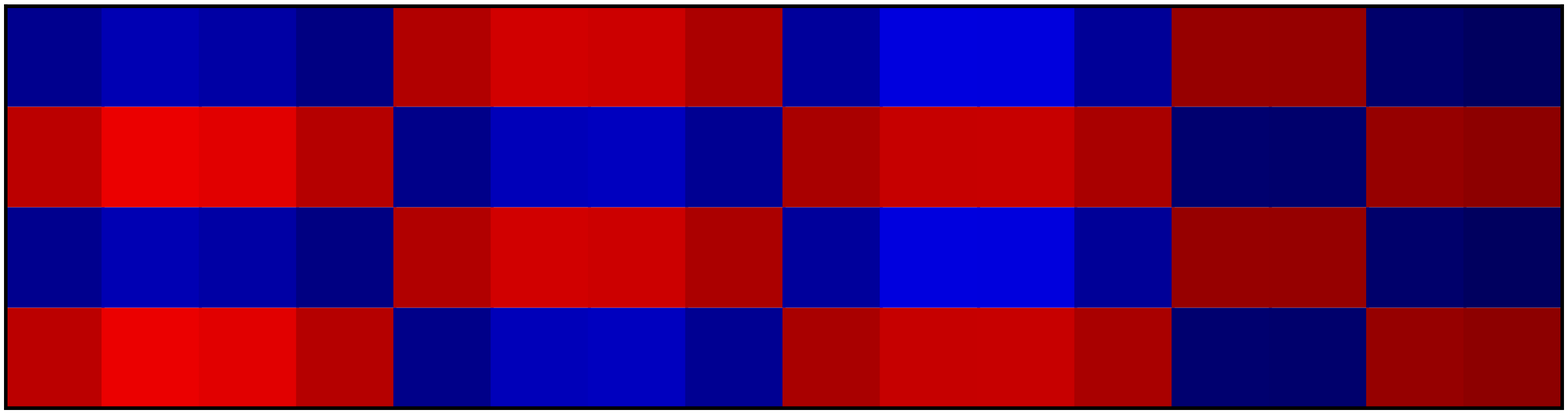}
 	\includegraphics[height=0.65cm, trim=0cm 0cm 0 3.2cm]{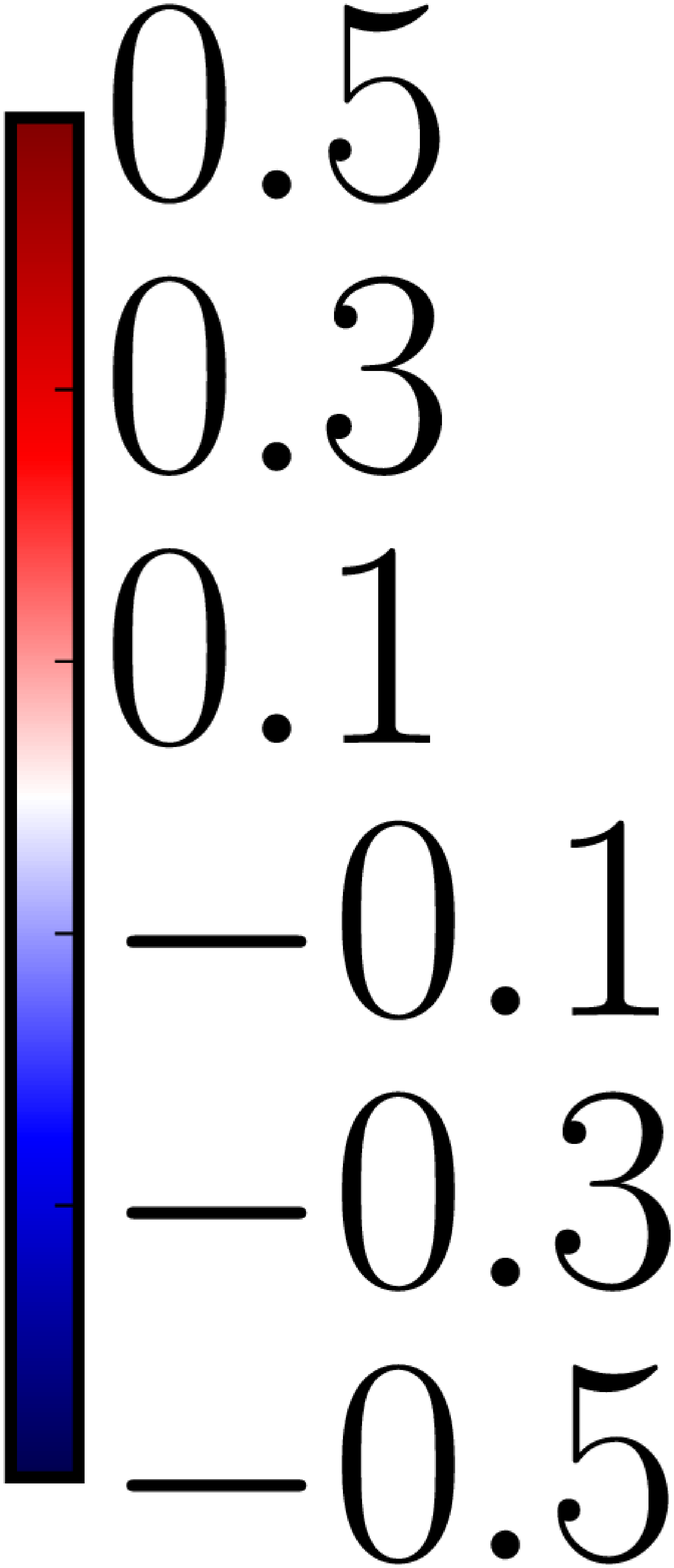}
 	\caption{Finding the optimal $U_{\rm eff}$ for the IP calculations during the iterations.
	The staggered spin density from a UHF calculation with $U_{\rm eff}$  is shown during
	each iteration of the self-consistent procedure with CPMC.
 	The self-consistency procedure begins with the free-electron $|\Psi_T\rangle$ in panel (a) 
		and with the $U=8t$ UHF solution in (b). In the lower panel, the left, middle and right depict
		 the spin density of the free electron wave function, the correct ground state from the converged
		 CPMC result, and the UHF solution with $U = 8t$, respectively.
		The system is the same as in Fig.~\ref{spin_4_16_mf}.
 		}
 	\label{spin_UHF}
 \end{figure}

We next examine the IP solutions during the self-consistent process. 
In Fig.~\ref{spin_UHF} we plot the staggered spin density from UHF, using the $U_{\rm eff}$ 
values which emerge during 
the iteration with CPMC calculations. 
(Each IP result is from the fully self-consistent UHF solution \cite{xu_jpcm_2011} at 
the indicated $U_{\rm eff}$.)
 The spin densities in the two initial trial wave-functions are very
  different, and neither gives the correct magnetic order, as shown by the patterns in the bottom row.
  With the iteration, the spin densities from the IP solutions become better. 
  At convergence, with a $U_{\rm eff}$ value of $~2.7t$ for this system,
  the densities from
  the UHF solution  are in fact quite close to the exact results. The reduction 
  from the ``bare" 
  $U$ of $8t$ is substantial, because of the tendency in UHF to severely over-estimate
 order. The self-consistent procedure allows an \emph{ab initio}
  determination of an optimum effective $U$.

\begin{figure}[t]
  	\includegraphics[width=8.0cm]{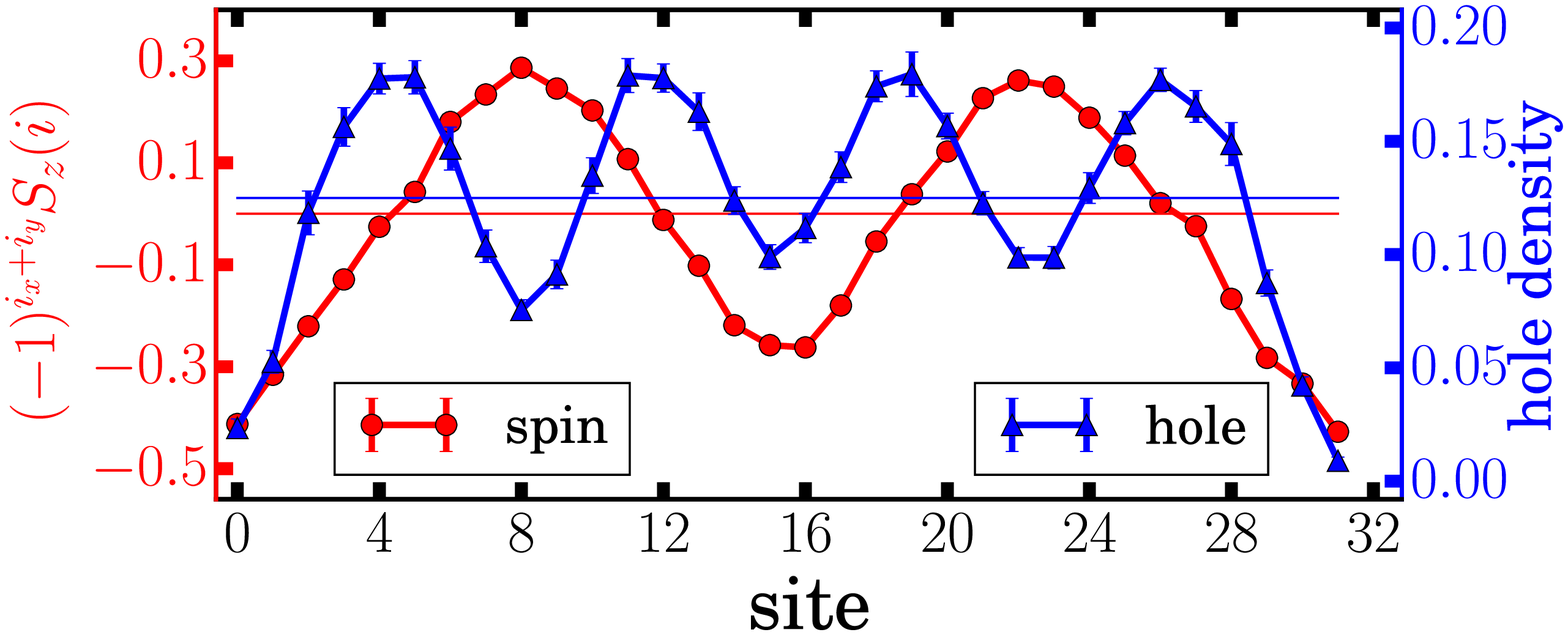}	
  	\includegraphics[width=4.5cm]{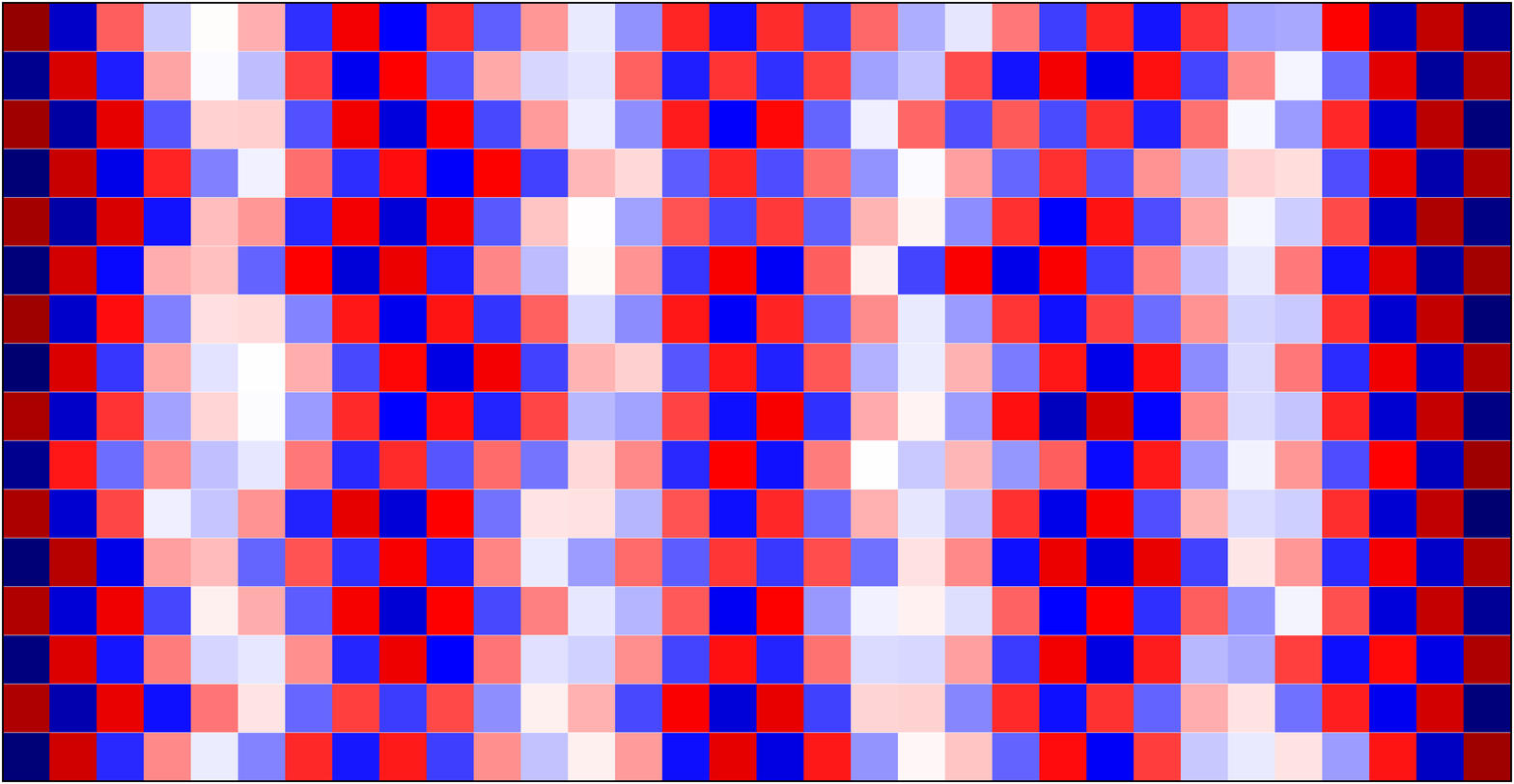}
  	\includegraphics[height=2.0cm, trim=0cm 0cm 0 0.2cm]{legend_color.eps}
  	\caption{ Converged CPMC results after self-consistent procedure for a large system, of $16 \times 32$.
  		In the upper panel, the staggered spin and hole densities are plotted. 
  		The red and blue horizontal lines represent zero spin-density and the average 
  		hole density, respectively. In the lower pane, the spin density for the cell is shown
  		with a colormap.
		As in the earlier systems,  $U = 8t$, $h=1/8$, and pinning field is applied to both edges along $L_y$. 
	}
  	\label{16_32_self}
  \end{figure}

Much larger system sizes must be treated 
in order to determine the nature of the magnetic order in the thermodynamic limit. 
The effect of the pinning fields must be minimized, $L_x$ needs to be sufficiently large to move from 
ladders to two-dimensions,   
and $L_y$ must be sufficiently large to 
accommodate the
wavelength of possible collective modes. 
This can now be achieved by 
the QMC self-consistent procedure. 
In Fig.~\ref{16_32_self} we show the results for a $16 \times 32$ lattice, with the same physical parameters.
The converged CPMC staggered spin and charge densities are plotted in the upper panel. The result 
confirms the tendencies of the
spin and charge orders seen in the smaller system sizes. 
A "bulk" region is present in the middle which gives a characteristic wavelength.
The
lower panel illustrates the spin-density wave structure, with the four 
nodal lines of modulation clearly visible (where the holes are more concentrated).
This is consistent with a wavelength of  $1/h$ seen at lower interaction strengths \cite{chia-chen_prl}.

\section{Discussion}

We have also tested a
different but related approach for constructing the trial wave function 
self-consistently from QMC.  
To encode the information on the one-body density matrix from CPMC, $\rho^{\rm CPMC}$, in the next stage $|\Psi_T\rangle$,
we seek a Slater determinant which gives a one body density matrix closest to $\rho^{\rm CPMC}$. 
This is done using 
natural orbitals, 
i.e., by diagonalizing:
\begin{equation}
\rho^{\rm CPMC}=
V\Lambda V^{\dagger}\,.
\end{equation}
The eigenvectors (natural orbitals) in $V$ corresponding to the $N_\sigma$ largest eigenvalues in $\Lambda $ are chosen to 
construct a Slater determinant.
This procedure is implemented for 
each spin specie $\sigma$  separately, 
leading to a UHF-like $|\Psi_T\rangle$. We found that this procedure gave  results similar
to the self-consistent approach in the systems tested above.

Although we have used the Hubbard model as an illustration, the self-consistent procedure we have proposed 
can be generalized to AFQMC calculations in real materials \cite{real_mat_self}, and
opens new directions to further improve the predictive 
power of calculations in correlated electron systems. In real materials, the self-consistent 
iteration with IP calculations can be used to improve an exchange-correlation functional, for example
to tune the optimal mixing parameter in a hybrid functional \cite{hybrid_fun_1,hybrid_fun_2}. The procedure could also be 
used to find a correct ``$U$" parameter in the context of LDA+$U$ \cite{LDA_U}. 
The idea of introducing an effective ``$U$" 
can also be connected to embedding theories \cite{DMET,DMFT},
although here the feedback of $U_{\rm eff}$ to the real system (cluster) is less direct.

\section{Summary}

In summary, we have developed a self-consistent constrained-path AFQMC method which allows
the QMC calculation to systematically improve its accuracy, while fully controlling the fermion sign or phase
problem. The paradigm coupling QMC with an IP calculation  allows a feedback from 
the fomer into the latter. 
This provides not only a way to improve the constraining trial wave functions for the (next iteration) QMC,
but also an independent-particle framework which in itself gives a drastically improved description 
of the physical system. 
The approach can be applied to 
strongly correlated models in
	condensed matter, ultra-cold atoms and optical lattices, nuclear shell models,
	as well as \emph{ab initio} calculations in molecules and solids. 

\begin{acknowledgments}
We are very grateful to Steven R. White and Chia-Min Chung for providing the DMRG results
and for many helpful communications. We acknowledge  support from NSF (DMR-1409510). 
MQ and SZ were also supported by the Simons Foundation.
The calculations were carried out at
the Extreme Science and Engineering Discovery Environment
(XSEDE), which is supported by National Science Foundation grant number ACI-1053575,
and the computational facilities at the College of William and Mary.
\end{acknowledgments}


\begin{thebibliography}{1}
	
	
\bibitem{QMC_prd_1981}
R.~Blankenbecler, D.~J.~Scalapino and R.~L.~Sugar, Phys. Rev. D {\bf 24}, 2278 (1981).

\bibitem{QMC_AP_1986}
G. Sugiyama and S. E. Koonin, Ann. Phys. (N.Y.) {\bf 168}, 1 (1986).

\bibitem{QMC_prb_1989}
S.~R.~White, D.~J.~Scalapino, R.~L.~Sugar, E.~Y.~Loh, J.~E.~Gubernatis, and R.~T.~Scalettar, Phys. Rev. B {\bf 40}, 506 (1989).
 
\bibitem{QMC_rmp_2011}
E. Gull, A. J. Millis, A. I. Lichtenstein, A. N. Rubtsov, M. Troyer, and P. Werner, Rev. Mod. Phys {\bf 83}, 349 (2011).

\bibitem{Hirsch_prb_1985}
J.~E.~Hirsch, Phys. Rev. B {\bf 31}, 4403 (1985).

\bibitem{Wu_prb_2005}
Congjun Wu and Shou-Cheng Zhang, Phys. Rev. B {\bf 71}, 155115 (2005).

\bibitem{sign_1}
E.~Y.~Loh Jr., J.~E.~Gubernatis, R.~T.~Scalettar, S.~R.~White, D.~J.~Scalapino, and R.~L.~Sugar, Phys. Rev. B {\bf 41}, 9301 (1990).

\bibitem{sign_2}
K. E. Schmidt and M. H. Kalos, in Applications of the Monte Carlo Method in Statistical Physics, edited by K. Binder (Springer-Verlag, Heidelberg, 1984).


\bibitem{GF_fermion}
D.~F.~B.~ten Haaf, H.~J.~M.~van Bemmel, J.~M.~J.~van Leeuwen, W.~van Saarloos, and D.~M.~Ceperley,
Phys. Rev. B {\bf 51}, 13039 (1995).



\bibitem{paper_simons}
J. P. F. LeBlanc, Andrey E. Antipov, Federico Becca, Ireneusz W. Bulik, Garnet Kin-Lic Chan, Chia-Min Chung, Youjin Deng, Michel Ferrero, Thomas M. Henderson, Carlos A. Jimenez-Hoyos, E. Kozik, Xuan-Wen Liu,
Andrew J. Millis, N. V. Prokofev, Mingpu Qin, Gustavo E. Scuseria, Hao Shi, B. V. Svistunov, Luca F. Tocchio, I. S. Tupitsyn, Steven R. White, Shiwei Zhang, Bo-Xiao Zheng, Zhenyue Zhu, and Emanuel Gull,
Phys. Rev. X {\bf 5}, 041041(2015).

\bibitem{DMC_RMP}
W.~M.~C. Foulkes,~L.~Mitas, R.~J.~Needs, and G.~Rajagopal,
Rev. Mod. Phys. {\bf 73}, 33 (2001);
J. Kolorenc and L. Mitas,   Rep. Prog. Phys. {\bf 74}, 026502 (2011);
L. K. Wagner and D. M. Ceperley, 
Rep. Prog. Phys.  {\bf 79}, 094501 (2016).

\bibitem{DMC_Devaux} See, e.g., 
N.~Devaux, M.~Casula, F.~Decremps, and S.~Sorella,
Phys. Rev. B {\bf 91}, 081101 (2015);
Kateryna Foyevtsova, Jaron T. Krogel, Jeongnim Kim, P. R. C. Kent, Elbio Dagotto, and Fernando A. Reboredo,
Phys. Rev. X {\bf 4}, 031003 (2014).

\bibitem{AFQMC-solids}F.~Ma, W.~Purwanto, S.~Zhang, H.~Krakauer, 
Phys. Rev. Lett. {\bf 114}, 226401 (2015).

\bibitem{CP_nuclear}
J.~Carlson, S.~Gandolfi, F.~Pederiva, Steven C.~Pieper, R.~Schiavilla, K.~E.~Schmidt, and R.~B.~Wiringa
Rev. Mod. Phys. {\bf 87}, 1067 (2015).

\bibitem{DMC_QC_1}
Hammond B L, Lester W A Jr and Reynolds P J, Monte Carlo Methods in Ab Initio Quantum Chemistry (World Scientific, Singapore, 1994).

\bibitem{DMC_QC_2} See, e.g., 
B. K. Clark, M. A. Morales, J. McMinis, J. Kim, and G. E. Scuseria,   J. Chem. Phys. {\bf 135}, 244105 (2011);
H. Zulfikri, C.~Amovilli, and Claudia Filippi, J. Chem. Theory Comput.,{\bf 12},1157  (2016).

\bibitem{AFQMC-QC}W.~A.~Al-Saidi, Shiwei Zhang and Henry Krakauer, J. Chem. Phys. {\bf 124}, 224101 (2006).

\bibitem{diff_self}
Several flavors of QMC exist which improve the result by bringing back the exponential computational
scaling
--- see, e.g.,
George H.~Booth, Alex J.~W.~Thom and Ali Alavi, J. Chem. Phys. {\bf 131}, 054106 (2009); 
F.~A.~Reboredo, R.~Q.~Hood, and P.~R.~C.~Kent, Phys. Rev. B {\bf 79}, 195117 (2009);
D.~M.~Ceperley and B.~J.~Alder, 
Phys.~Rev.~Lett.~{\bf 45}, 566 (1980);
and Ref.~\cite{CPMC_sym_1}.
We focus here on methods that retain low-algebraic computational scaling to treat extended systems.

\bibitem{VMC}
C.~J.~Umrigar, K.~G.~Wilson, and J.~W.~Wilkins, Phys. Rev. Lett. {\bf 60}, 1719 (1988);
C. J. Umrigar, Julien Toulouse, Claudia Filippi, S. Sorella, and R. G. Hennig
Phys. Rev. Lett. {\bf 98}, 110201 (2007).

\bibitem{Gilbert_1975}
T. L. Gilbert, Phys. Rev. B {\bf 12}, 2111 (1975).

\bibitem{lecture-notes}
S. Zhang, Auxiliary-Field Quantum Monte Carlo for Correlated Electron Systems, Vol. 3 of Emergent Phenomena in Correlated Matter: Modeling and Simulation, Ed. E. Pavarini, E. Koch, and U. Schollwock (Verlag des Forschungszentrum Julich, 2013).

\bibitem{zhang_prb_1997}
S.~Zhang, J.~Carlson, and J.~E.~Gubernatis, Phys. Rev. B {\bf 55}, 7464 (1997).

\bibitem{zhang_prl_2003}
S.~Zhang and H.~Krakauer, Phys. Rev. Lett. {\bf 90}, 136401 (2003).


\bibitem{chia-chen_prb}
Chia-Chen Chang and Shiwei Zhang, Phys. Rev. B {\bf 78}, 165101 (2008).


\bibitem{CPMC_sym_1}
Hao Shi and Shiwei Zhang, Phys. Rev. B {\bf 88}, 125132 (2013).

\bibitem{CPMC_sym_2}
Hao Shi, Carlos A.~Jimenez-Hoyos, R.~Rodriguez-Guzman, Gustavo E.~Scuseria, and Shiwei Zhang, Phys. Rev. B {\bf 89}, 125129 (2014).

\bibitem{hubbard_benchmark}
Mingpu Qin, Hao Shi, and Shiwei Zhang, Phys. Rev. B {\bf 94}, 085103 (2016).


\bibitem{Wirawan-PRE}
Wirawan Purwanto and Shiwei Zhang, Phys. Rev. E {\bf 70}, 056702 (2004).


\bibitem{stripe_exp}
J. M. Tranquada, B. J. Sternlieb, J. D. Axe, Y. Nakamura, and S. Uchida, Nature {\bf 375}, 561 (1995).

\bibitem{sr_white_dmrg_original}
S. R. White, Phys. Rev. Lett. {\bf 69}, 2863 (1992),  S. R. White, Phys. Rev. B {\bf 48}, 10345 (1993).

\bibitem{srwhite_prl_2007}
Steven R. White and A. L. Chernyshev, Phys. Rev. Lett. {\bf 99}, 127004 (2007).

\bibitem{srwhite_prb_2009}
Steven R. White and D. J. Scalapino, Phys. Rev. B {\bf 79}, 220504 (2009).

\bibitem{assaad_prx_2013}
Fakher F. Assaad, and Igor F. Herbut, Phys. Rev. X {\bf 3}, 031010 (2013).

\bibitem{xu_jpcm_2011}
Jie Xu, Chia-Chen Chang, Eric J. Walter, Shiwei Zhang, J. Phys.: Condens. Matter {\bf 23}, 505601 (2011).

\bibitem{diff_row}
For hole density, the results are the same for each row, while for spin density, there is a $\pi$ phase between
even and odd rows.


\bibitem{Carlson-PRB1999}
J.~Carlson, J.~E.~Gubernatis, G.~Ortiz, and Shiwei Zhang, Phys. Rev. B {\bf 59}, 12788 (1999).

\bibitem{chia-chen_prl}
C.-C.~Chang and S.~Zhang, Phys. Rev. Lett. {\bf 104}, 116402 (2010).

\bibitem{real_mat_self} Mario Motta, Shiwei Zhang, et al., to be published.

\bibitem{hybrid_fun_1}
R.~M.~Martin, Electronic Structure (Cambridge University Press, Cambridge, 2004).

\bibitem{hybrid_fun_2}
John P.~Perdew, Matthias Ernzerhof and Kieron Burke, J. Chem. Phys. {\bf 105}, 9982 (1996).

\bibitem{LDA_U}
Vladimir I. Anisimov, Jan Zaanen, and Ole K. Andersen, Phys. Rev. B {\bf 44}, 943 (1991).
 
\bibitem{DMET}
Gerald Knizia and Garnet Kin-Lic Chan, Phys. Rev. Lett. {\bf 109}, 186404 (2012).

\bibitem{DMFT}
Antoine Georges, Gabriel Kotliar, Werner Krauth, and Marcelo J. Rozenberg
Rev. Mod. Phys. {\bf 68}, 13 (1996).

	
\end{thebibliography}
\end{document}